\documentclass[aps,prd,showpacs,twocolumn]{revtex4}

\usepackage{latexsym}
\usepackage{amsfonts,amssymb}
\usepackage{graphicx,epsfig}
\usepackage{psfrag}

\newcommand{\be}{\begin{equation}}
\newcommand{\ee}{\end{equation}}
\def\bq{\begin{eqnarray}}
\def\eq{\end{eqnarray}}
\def\n{\nonumber}
\def\t{\tau}

\def\ep{\epsilon}

\def\a{\alpha}
\def\th{\theta}

\def\s{\sigma}

\def\La{\Lambda_4}
\def\Laf{\Lambda_4}

\begin{document}

\title{Localization of gravity in brane world cosmologies}

\author{Parampreet Singh\footnote{e-mail: param@iucaa.ernet.in} and
Naresh 
Dadhich\footnote{e-mail: nkd@iucaa.ernet.in }}
\address{Inter-University Centre for Astronomy and Astrophysics,\\
Post Bag 4, Ganeshkhind, Pune-411 007, INDIA.}


\begin{abstract}                                    
The most remarkable and interesting feature of brane world scenario is
the use of bulk's curvature 
to localize gravity on the brane, albeit with fine tuning of the brane
and bulk parameters. For FRW expanding universe on the brane, it is a
moving hypersurface in Schwarzschild anti de Sitter bulk spacetime. We show
that zero mass gravitons have bound state on the brane for suitable values of
brane and bulk parameters. There occur various cases giving rise to different
cosmological models, in particular we discuss a model with positive 
cosmological constant on the brane.
\end{abstract} 

 
\maketitle       


The view that our Universe might actually have dimensions more than
four is anchored on the recent developments in string and M-theories in
which gravity arises as a truly higher dimensional interaction. Only in the
low energy limit it manifests in the familiar $4$-dimensional general
relativity (GR). This has initiated a lot of activity in recent times. 
Though the basic idea was already  there in the form of Kaluza-Klein (KK) theories, the recent spurt is
primarily due to the possibility which helps solve the mass hierarchy
problem in the  standard model of particle physics. 

In some of these models \cite{arkani} the extra dimensions 
can be as large as millimeter which
 is however less than the current observational limits on low scale
gravity.
Notable are the models in which the extra dimensions are allowed to be
of infinite extent \cite{rs1,rs2}. These models have $Z_2$-symmetry, which
is motivated by the reduction of M theory to $E_8 \times E_8$ heterotic
string theory \cite{witten}. The single brane Randall-Sundrum (RS) 
model \cite{rs2} has attracted a lot of interest and activity.
In this model the Minkowski flat brane in $5$-D anti de Sitter (AdS)
bulk has a positive tension. It is then possible to recover Newton's inverse
square law with $r^{-4}$ correction term which arises from massive KK modes
contribution. 
There have been various generalizations of this \cite{dick} in the 
form of thick branes \cite{csaki}, AdS branes \cite{karch}
and brane models without $Z_2$ symmetry \cite{colyuri}.

In the overall view of the brane world scenario, all matter fields live
on the 3-brane which is the $4$-D physical Universe while gravity can
propagate in the extra dimensions, say 5-D bulk. The bulk 
and brane spacetimes are joined together through the Israel junction
conditions \cite{israel}. Since, the standard Einstein 
equations on the brane are
modified by the bulk effects it opens a whole new vista for
investigation of 
astrophysical and cosmological consequences of these models (see for
eg. \cite{cosmo1}). The connection with
CFT correspondence has also been studied (see for eg. \cite{cft}). For
complete solution of the 
problem, one has to solve the $\Lambda$-vacuum equation in $5$-D for
the bulk spacetime and simultaneously the modified Einstein equation which in
addition to the square of stress tensor also 
contains the projection of the bulk
Weyl curvature on the brane (dark radiation) and then the two solutions
should be joined together with the Israel junction conditions \cite{israel}.
It is by all means a very formidable task and it is therefore no surprise
that there exist only few complete solutions to date. The two important
cases are flat/vacuum brane with AdS bulk and FRW brane with
Schwarzschild-AdS (S-AdS) bulk \cite{bdel,kraus,s-ads}.

The S-AdS bulk-brane system is composed of two patches of S-AdS bulk having in
general different mass parameters for the black hole with the brane located
at $y = y(\t)$, where $y(\t)$ is determined by solving the Israel junction conditions \cite{kraus}. From the resulting
equation for brane trajectory  one finds
that in  S-AdS bulk, the FRW brane will in general be moving unless the
parameters are 
properly fine tuned (eg. for the RS case $\s = (3/ 4 \pi G_5 l),
\Lambda_4 = 0$, where $\s$ is the brane tension, $G_5$ is the $5$-D 
gravitational 
constant and $l$ is the radius of curvature  of the bulk spacetime).
 The extra dimension is a radial coordinate
of the bulk and imposing $Z_2$ symmetry across the brane demands that 
the mass parameters
of both the patches to be same. By fine tuning parameters one can obtain
static branes which cannot harbour expansion which is essential for realistic
cosmological models. A slight off tuned value of $\s$ or non 
zero value of black hole mass would set it moving.

Though localization of gravity for the AdS bulk with flat brane and for
some generalizations of it has been established 
\cite{karch,langlois2,odintsov}, it has not yet been investigated for the S-AdS bulk with FRW
brane. The problem becomes difficult if one notes that unlike the RS case, the brane is dynamic in the static bulk and it is non trivial to fix the boundary conditions on the modes.  
For localizability of gravity on the brane, first there should exist bound
normalizable mode for zero mass graviton and plus there should also exist
KK modes to give the correction to the Newtonian gravity. All this could be
studied by considering perturbation of the bulk metric and the brane motion determined by the Israel junction conditions.
This is what we shall concern 
ourselves in this investigation which should be quite indicative of
localization of gravity on the brane. For actual demonstration of recovering
Newtonian gravity with the correction, one has to do propagator analysis
which is hard to carry out for a non-static brane and we shall not
attempt that here.

It is well known that localizability is very sensitive to fine
tuning of parameters. A priori, there is no well established criterion to
check this. For instance there does exist a bulk spacetime, which is an exact
solution of the $\Lambda$-vacuum equation, for which gravity is non localizable on the brane \cite{nkd1}. This is the case of Nariai metric which has non zero
Weyl curvature. Note that Weyl is non zero for S-AdS as well. It is
therefore pertinent to find under what conditions do zero mass gravitons have
bound state on FRW brane? This is the most critical question for brane world
cosmologies and that is what we wish to address in this letter.

Since our brane is a hypersurface around the black 
hole, any movement of the brane towards or away from the black hole
would be interpreted as contraction or expansion by the observers on
the brane. 
That is how expansion is directly related to motion of the
brane. 
To see whether zero mass graviton has a bound normalizable state in these cosmological models one has to solve for the graviton perturbation equation by
taking into account the location of the brane. 
Unlike the static cases (like the RS case) the position of the FRW brane
would not be fixed in the bulk and  one would have to take 
into account the trajectory of the brane to understand the localizability. 
The brane trajectory found through 
Israel matching conditions would be given by a Friedmann like equation
which determines the position of the brane. In the following we shall hence
study the metric perturbations of the S-AdS bulk and obtain a potential 
in the graviton wave equation which would determine the fate of localizability
once the brane trajectory (and hence the location of the brane) is taken care
of. We shall then show how the RS case can be recovered in this scenario,
then we shall in particular consider the case
of $k = 1, \La > 0$, while a comprehensive discussion of all possible
cases would be done elsewhere \cite{next}.

In the five dimensional bulk we have the S-AdS metric, 
\be
d s^2 = - e^{2 \beta} \, d t^2 + e^{- 2 \beta} d y^2 +
y^2 \bigg[ \frac{d r^2}{1 - k r^2} + r^2 d \Omega_2^2
\bigg] \label{eq:metric}
\ee
where 
\be
e^{2 \beta} = \left( k + \frac{y^2}{l^2} - \frac{M}{y^2} \right).
\ee
Here $k = 0,\pm1$ is the curvature index and $M$ 
is the mass parameter of the bulk black hole. 

The static limit  would be 
given by $g_{00} = 0$ which leads to,
\be
y_h^2 = \frac{l^2}{2} \, \left(- k + \sqrt{k^2 + \frac{4 M}{l^2}}
\right)
\label{eq:horizon}
 \ee
and since the metric is spherically symmetric this also defines the 
location of event 
horizon of the spacetime. The above metric is the solution
of the 
$\Lambda$-vacuum equation,
\be
G_{ab} = - \Lambda_5 \, g_{ab}, \hspace{1cm} \Lambda_5 = - 6/l^2.
\ee
The Latin indices which label the bulk spacetime $(x^\mu, y)$ run from 0...4 and the Greek indices labelling brane spacetime $(x^\mu$) would run from
0...3. 
We consider the metric perturbations of the above metric
$g_{ab}^{(B)}$, 
i.e. $h_{ab} = g_{ab} - g_{ab}^{(B)}$. We would take the metric
perturbations in the extra dimension to vanish, $h_{t \, y} = h_{r \,
y} = h_{\th \, y} = h_{\phi \, y} = h_{y \, y} = 0$. We would further impose
the transverse-traceless gauge conditions, $\nabla^{\mu} h_{\mu \nu} = 0, 
\, h^{\mu}{}_{\mu} = 0$ and hence the wave equation turns out to be,
\be
\Box_{5} h_{ab} + 2 \, R^{(B)}_{cadb} \, h^{cd} - R^{(B)}_{a c} \, h_b{}^{\, c} -
 R^{(B)}_{b c} \, h_a{}^{\, c} 
= \Lambda_5 \, h_{ab}.\label{eq:fwaveq}
\ee

Assuming that  wavelength of the 
gravitons is much smaller than the radius of curvature of the background
and their amplitude is very small we can work under linearized
approximation. We can further choose a constant vector field which
satisfies $h_{a b} u^b = 0$ and hence we are effectively left with two
independent modes. 
We solve this wave equation with the ansatz that 
$h_{a b} (x^\mu, y) = h_{ab} (x^\mu) \, \Psi(y)$. Substituting this
in eq.(\ref{eq:fwaveq}) and using  $m^2$ as the constant of separation of variables, the $y$ dependence turns out to be
\be
\left(\frac{y^2}{l^2} - \frac{M}{y^2} + k \right) \Psi'' + \left(\frac{3 M}{y^3} - \frac{k}{y} + \frac{y}{l^2} \right) \Psi' - \frac{4 M}{y^4} \Psi + m^2 \Psi = 0  \label{eq:y}
\ee
where prime denotes a derivative with respect to $y$. This equation can be written down in the form 
\be
\Psi'' + a_1(y) \Psi' + a_2(y) \Psi = 0
\ee
which can be transformed into a wave equation form with the transformation $\Psi(y) = \phi(y) \psi(y)$
where 
\be
\phi(y) = c \, exp \left( -\frac{1}{2} \, \int_{y_i}^{y_f} \, a_1(y) \, dy \right).
\ee
Here $c$ is a constant and  $y_i$ and $y_f$ belong to the interval over which
$a_1, a_2$ are continuous and $a_1$ has a continuous derivative. This transformation  eliminates the first order derivative term in eq.(\ref{eq:y}) 
and we get the desired  Schroedinger like equation
\be
- \frac{1}{2} \psi(y)'' + V \psi(y) = m^2 \psi(y) \label{eq:yeq} .
\ee
The potential consists of two parts, $V = V_f + V_m$, 
which are given by   
\bq
V_f =  &-& \n
 \frac{6 l^2 (5 M y^4 - k y^6)}{8 (y^5 + l^2(- M y + k y^3))^2} 
+ \frac{2 M l^2}{y^2 (y^4 - M l^2 + k l^2 y^2)} \\
 &-&  \frac{y^8 + l^4( - 15 M^2 + 22 k M y^2 - 3 k^2 y^4)}{8 (y^5 +
l^2(- M y + k y^3))^2} \label{eq:vf}
\eq
and the interaction part
\be
V_m = - \frac{m^2 l^2 y^2 - 2 m^2(y^4 - M l^2 + k l^2 y^2)}{2(y^4 - M
l^2 + k l^2 y^2)} . \label{eq:vm}
\ee   
 The presence of interaction
part is peculiar to the coordinate system we are working with.
 Obviously,
the shape of the potential will play determining role in localization of 
gravitons. The potential  $V$ for all the cases provides a negative
infinite well at the event horizon of the black hole of mass $M$,
which holds irrespective of the matter distribution on the brane. The potential
for the massless and massive modes for the case $k = 1, M \neq 0$ is shown
in Fig. 1.  The potential for the massless mode can be viewed as the $m \rightarrow 0$ limit of the potential for massive modes for which $V > 0$ for 
large $y$. Such a behaviour is a generic feature of various cosmological
scenarios.

 The Israel junction conditions determine the relationship
between the bulk and brane parameters, 
\be
\Lambda_4 = \frac{\Lambda_5}{2} + \frac{16 \pi^2}{3} \, G_5^2 \sigma^2,
\, 
G_4 = \frac{4 \pi}{3} \, G_5^2 \, \s
\ee
and also the motion of the brane given by the equation  \cite{kraus},
\be
{\dot y}^2 - \frac{8 \pi G_4}{3} \, \rho \,\left(1 + \frac{\rho}{2 \s}
\right) y^2 + (1 - \s^2/\s_0^2) \, \frac{y^2}{l^2} -  \frac{M}
{y^2}  + k = 0. \label{eq:bdynawm}
\ee
Here dot refers to derivative relative to proper time $\t$ and
$\rho$ is 
the energy density on the brane and the critical tension $\s_0 \equiv 
 3/4 \pi G_5 l$, which determines the sign of $\Lambda_4 $. 
The $y$ coordinate now plays a dual role.
It not only tells us about the effect on graviton fluctuations due to extra dimension but also parameterizes the brane trajectory. The metric on the
brane is FRW and is given by,
\be
d s_4^2 = - d\t^2 +  y^2(\t) \left(\frac{d r^2}{1 - k r^2} + d \Omega_2^2
\right)
\ee   
where  $d\t^2 = exp(2 \beta(t))(1 - exp(- 4 \beta(t)) (d y/ d t)^2 ) d
t^2 $. The physically interesting branes would lie outside the horizon which would also serve as the high
energy cutoff. For localization of massless mode we require it to
be bounded as well as normalizable. The form of the potential for
various choice of cosmological parameters is such that boundedness
of massless mode is guaranteed once the brane crosses horizon. However,
the normalizability
of the massless mode depends on the asymptotic behavior of the potential and
the form of the potential  reflects that branes which are either static or ever expanding in the bulk would harbour localization. 

As is clear from eq.
(\ref{eq:bdynawm}) 
that motion of brane is determined by both
black hole mass and energy distribution on the brane.
The localization would therefore depend on energy distribution in both
bulk and brane. The localization of the massless
mode depends on whether the brane expands forever or has a bounce,
and it can be easily checked from eq.(\ref{eq:bdynawm}) that 
such a behaviour of the brane trajectory is unchanged even if the
matter density terms are excluded. For simplicity we would henceforth
exclude matter density terms from our analysis of brane trajectory
and hence the brane trajectory equation becomes
\be
{\dot y}^2 - \frac{\Lambda_4}{3}  y^2  -  \frac{M}{y^2}  + k= 0
\label{eq:bdyna} .
\ee
 Note that this equation is non linear
in the
highest order of derivative and hence it will not have unique
solution.

It should be noted for  dynamic branes the proper time as measured by the observers on the brane is not the same as the time component of the bulk metric.
For the observer on the brane to interpret the massless mode correctly
one thus has to assume that  $y(t)$ is a slowly varying function. Further,
fixing boundary conditions for the modes on a moving boundary is a very
difficult task. However, in a cosmological setting of FRW branes we can
safely assume that brane is slowly expanding and these problems can be
bypassed. Our results for dynamic branes would hence hold in the above cosmological scenario.

\begin{figure}[tbh!]
\epsfig{figure=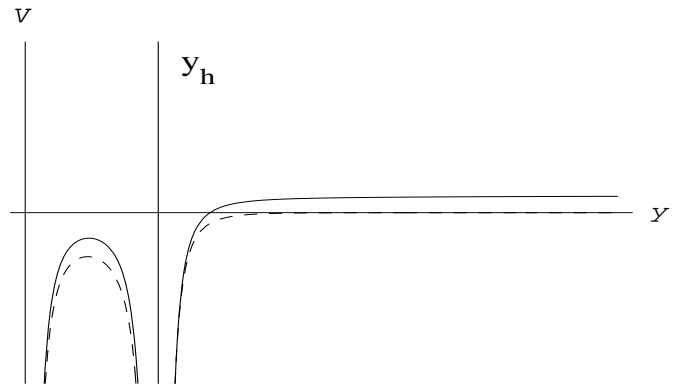,height=2in,width=3.5in,angle=0}
{\caption{\small Potential plot for $M \neq 0$ and  $k = 1$. The dashed and dark lines indicate potential
for  massless 
 and massive mode respectively.}}
\end{figure}

Now we would see how the above metric perturbation equation and the
brane 
dynamics equation yield RS model. For this $k = M = 0, \s = \s_0$, the
metric 
eq.(\ref{eq:metric}) takes the form
\be
d s^2 = -\frac{y^2}{l^2} d t^2 + \frac{l^2}{y^2} d y^2 + y^2 \left(d r^2 + r^2 d r^2 + r^2 \, \sin^2\th d \phi^2
\right) .  
\ee
Using the transformation $\eta = l \, \ln(l/y)$ one can rewrite this metric
in the familiar RS form. Note that $y$ is a radial coordinate of the black hole
in the bulk and it ranges from $0$ to $\infty$. The above transformation 
would yield $\eta$ ranging from $- \infty$ to $\infty$. Imposition of
$Z_2$ symmetry across the brane (i.e. matching the extrinsic curvatures
on both sides of the brane) does not give any condition on $y$ coordinate,
however in $\eta$ coordinate it demands   $\eta_{(-)}$ to be identified
with $\eta_{(+)}$ and hence the metric  contains $|\eta|$ in the warp 
factor,
\be
d s^2 = d \eta^2 + e^{- 2 |\eta|/ l} (- dt^2 + dx_1^{2} + dx_2^{2} +
dx_3^{2}). 
\ee
 RS brane is static and located at $\eta = 0$, which means at $y = l$.
Note that the shape and form of the potential in the Schroedinger equation critically
 depends on the coordinates one is working with. In the $\eta$ coordinates 
it is the presence of $|\eta|$ which gives a Dirac delta potential at the
location of the brane. However, working with the $y$ coordinate one gets
from eqs.(\ref{eq:vf}) \& (\ref{eq:vm}), $V =  - 1/8y^2 + m^2 (1 - l^2/2y^2)$.
At the location of the brane the massless mode is bound and the
asymptotic behaviour of
the potential suggests that it is normalizable.
The massive modes for which $m > 1/2l$,
$V > 0$, are unbound. Solving near the brane one can  obtain the
wavefunctions from eq.(\ref{eq:yeq}) which turn out to be similar to 
those obtained by RS, with the required $m l$ suppression of
KK modes on the brane \cite{next}. The fact that 
$V$ turns positive for massive modes gives the continuum spectrum and
the RS correction to Newtonian potential. 
The form of the potential also suggests
that there exist discrete modes for  $m < 1/2l$. The appearance of 
discrete modes in this case is due to our choice of the coordinates. This is
clear from the fact that unlike RS coordinates our extra dimension is a 
radial coordinate in the bulk and our brane is curved. However, it should be
noted that the correction to Newtonian gravity is not expected to change
since the Green's function used by Garriga and  Tanaka \cite{gt} 
to evaluate the force law between two point sources on the brane in the 
RS case is inert under our transformation of coordinates from $y$ to $\eta$.
In the $y$ coordinates their Green's function can be written as
\bq
G_{R} (x, x') &=& \n - \int \, \frac{d ^4 k}{(2 \pi)^4} \, e^{i k_\mu ( x^\mu - {x^\prime}^{\mu})} \bigg[ \frac{(y^2/l^2) (y'^2/l^2)}{l ({\bf k}^2 - (w + i \ep)^2)} \\
&+&  \int_0^\infty \, \frac{u_m(y) u_m(y') ~ dm}{ m^2 + {\bf k}^2 - (w + i \ep)^2} \bigg]
\eq
where   $u_m(y)$ are the wavefunctions for massive modes. In the stationary    
case the Green's function between two points at the location of the brane,
i.e. $y = y' = l$ yields,
\be
G (x^i, l, {x^\prime}^{i}, l) \approx - \frac{1}{4 \pi l r} \left(1 + \frac{l^2}{2 r^2} \right)
\ee
which leads to the same $l^2/r^2$ correction to Newtonian gravity.

 For negative $\La$  branes the brane trajectory exhibits 
a bounce, like
for the case $k = -1, M = 0, \Lambda_4 < 0$ it is given by $y(\tau) =
\sin(\sqrt{-\Lambda_4/3} \tau)/\sqrt{-\Lambda_4/3} $. The potential is
similar
as in RS case, apart from that it blows up at origin and there is an
infinite
negative well at $y=l$, but the brane trajectory is such that it does
not
yield the required  behaviour of the potential for the
ground state
wavefunction  to be normalizable.
Hence the massless mode would not be localized for the negative $\La$
 brane (but that would be for the positive $\La$ brane because the 
solution of the brane motion equation
does permit normalized wavefunction). We thus recover the well-known
result for AdS bulk and negative/positive $\La$ branes  \cite{karch}. 
The behavior of
brane trajectory also plays critical role in localization of gravity on
the brane.

Now we turn to  FRW brane. Eq.(\ref{eq:bdyna}) yields host of
solutions
for interesting cosmological scenarios including the inflationary
solutions for all values of $k$ with a positive $\La$ on the brane,
which is 
also favored by current observations of type Ia supernovae \cite{sn}. 
A detailed discussion of all these cases would be done elsewhere
\cite{next}, 
here we would as a representative consider the case of $k = 1$, $M \neq
0$ 
and $\La > 0$.

 Solving eq.(\ref{eq:bdyna}) for this case  we get,
\be
y(\t) = \sqrt{\frac{3}{2 \La}} \, \bigg[ 1 + n \sinh(2 x) - \cosh(2 x)
\bigg]^{1/2} 
\ee
where $n = 2 \sqrt{M \Laf/3}$ and $x = \sqrt{\Laf/3} \tau$.
The behavior of potential $V$ for this case is shown in Fig.1. Note
that $V$ 
blows up  at $y = 0$ and there is an infinite well at $y = y_h$, we
shall 
therefore restrict to $y > y_h$. 
Since the form of the potential for localization requires an ever
expanding brane, only $n > 1$ is possible.
The brane  emerges out of the event horizon, 
expanding from $y = 0$ at $\t = 0$ like a white hole \cite{wh} and
would expand for ever, exponentially for large $\t$. The Hubble parameter 
for $k = 1$, $M \neq 0$ cosmological model is given by
\be
H = \frac{\dot y}{y} = \sqrt{\frac{\Laf}{3}} \, \frac{\bigg[ n \,
\cosh(2 x) - \sinh(2 x) \bigg]}{\bigg[ 1 + n \, \sinh(2 x) - \cosh(2 x)
\bigg]} \, \, 
 \ee
implying an inflationary universe at large $\t$.

For the massless mode the potential profile
favors boundedness and normalizability. The massive modes would be unbounded
and would contribute to correction over Newtonian potential. 
In order to show the correction suggested by  massive modes we would first
parameterize the location of the brane off the horizon by $\a = 1 - M/y^2
$.
 Solving the Schroedinger equation, eq.(\ref{eq:yeq}), in the approximation $M \ll l^2$ and near the horizon ($y_h \sim \sqrt{M}$) we get,
\be
\psi(y) =  \sqrt{y} \left( C_1 I_{-\gamma/2} (\nu \, y) + C_2 
I_{\gamma/2} (\nu \,  y) \right)
\ee  
where $\gamma=\sqrt{1 - 4 l^4(1 - 4 \a)(M - 2 \a l^2)/M^3}$
and $\nu=\sqrt{(1 - \a )( 2 \a l^2 - M)} m l /M$. 
The boundary conditions on the wavefunction at the brane in a cosmological
setting leads to $C_1 = (m l)^{(1 + \gamma)/2} M^{\gamma/4}$ with $\gamma < 2$ and $C_2
= 0$. Close to the horizon $\nu \sim m l/\sqrt{M}$ and since
$y \sim \sqrt{M}$, the argument in the Bessel function as in the RS case is proportional to
$m l$ at the location of the brane. This suggests a $l^2/r^2$ correction to Newtonian potential due to massive modes.

 The  modifications to the 
standard GR would be most prevelant near the event 
horizon which is the high energy end and marks a cut off for the
scale
factor $y(\tau)$. As the brane moves out and expands, the 
potential as 
shown in Fig.1 becomes shallower and the high energy modifications die
out 
with time.  In particular, the RS correction  to the Newtonian
potential 
will die out as the universe expands. 
 
The overall picture that emerges is that we could have different FRW 
cosmological models on the brane which are anchored onto S-AdS bulk.
The curvature of the bulk spacetime and the brane trajectory are key
players
in the localization of gravity  on the brane. The ultimate 
evolutionary fate of models is decided by the black 
hole mass and the cosmological constant on the brane.
We have thus shown that S-AdS bulk does allow localization of gravity
on FRW brane with $\La > 0$.

 In conclusion, we could say that brane cosmologies are firmly anchored
in respect of localization of gravity. We have shown that the potential in the
Schroedinger equation and the brane dynamics lead to bound state for  zero mass
graviton on a slowly moving brane in a cosmological setting.

We thank Roy Maartens and S. Shankaranarayanan for helpful discussions
and useful comments. Thanks are also due to colleagues who, in response to
earlier version, brought their relevant work to our notice. We also thank
Per Kraus for reaffirming on the issue of $Z_2$ symmetry in this work.
 PS thanks Council 
for Scientific \& Industrial Research for grant number: 2 -
34/98(ii)E.U-II.



\end{document}